\documentstyle[12pt]{article}
\newcommand{\beq}{\begin{equation}}
\newcommand{\eeq}{\end{equation}}

\def\p1half{{\textstyle{{{p+1}\over{2}}}}}

\def\23phalf{{\textstyle{{{23-p}\over{2}}}}}

\begin{document}
\thispagestyle{empty}
\begin{titlepage}

\bigskip
\hskip 3.7in{\vbox{\baselineskip12pt
}}

\bigskip\bigskip\bigskip\bigskip
\centerline{\large\bf
The Power of Worldsheets:}
\bigskip
\centerline{\large\bf
Applications and Prospects}
\bigskip\bigskip
\bigskip\bigskip
\centerline{\bf Shyamoli Chaudhuri
\footnote{Updated version of an invited talk given at the Tohwa International 
Symposium in String Theory, Fukuoka, Japan, July 2001.
Email: shyamoli@thphysed.org}
}
\centerline{214 North Allegheny St.}
\centerline{Bellefonte, PA 16823}
\date{\today}

\bigskip\bigskip
\begin{abstract}
We explain how perturbative string theory can be
viewed as an exactly renormalizable Weyl invariant quantum mechanics in
the worldsheet representation clarifying why string scattering amplitudes
are both finite and unambiguously normalized and explaining the origin of
UV-IR relations in spacetime. As applications we examine the worldsheet
representation of nonperturbative type IB states and of string solitons.  
We conclude with an analysis of the thermodynamics of a free closed string
gas establishing the absence of the Hagedorn phase transition. We show that 
the 10D heterotic strings share a stable finite temperature ground state with
gauge group $SO(16)$$\times$$SO(16)$. The free energy at the self-dual
Kosterlitz-Thouless phase transition is minimized with finite entropy and 
positive specific heat. The open and closed string gas transitions to a
confining long string phase at a temperature at or below the string scale
in the presence of an external electric field.
\end{abstract}

\end{titlepage}

\section{Introduction}

In this talk, I will make the argument that our strongest insights into
a fully nonperturbative framework for string/M theory 
are to be found in the worldsheet formalism for perturbative string
theory \cite{polyakov,poltorus}. On the other hand,
since the spacetime low energy effective action is the likely first point of
detailed contact with any candidate nonperturbative framework, it is crucial
that any wisdom gained from worldsheet calculations is incorporated into
formulating precise relations among both the couplings in the spacetime
effective action, as well as the {\em quantum amplitudes of the low energy
effective theory at the string scale}. Such relations may be interpreted
as the specification of renormalization conditions on the candidate
nonperturbative theory holding at the string mass scale, $\alpha^{\prime -1/2}$.
They will play an important role in the next wave of research into candidate
nonperturbative frameworks such as the reduced matrix models of
\cite{bfkt}.

\vskip 0.07in
Perturbative string theory originated in the Dual Model, a
phenomenological description of mesons and hadrons that pre-dates
the discovery of asymptotic freedom and subsequent development
of QCD as the theory of the strong interactions \cite{gsw,polbook}.
The spacetime viewpoint and introduction of the worldsheet formalism
came with the realization that many properties of dual
model amplitudes, such as $\rm S$$-$$\rm T$ channel duality,
are automatic in a quantum
theory of one-dimensional extended objects or {\em strings}.
The open string endpoints carry charge, and the lowest-lying excitation in the
open string spectrum is identified with a vector boson.
Nonabelian symmetry is straightforwardly introduced by assigning the
endpoint wavefunctions
to the fundamental representations of an internal symmetry group.
An added bonus is the discovery that the one-loop amplitude of an open
string theory always factorizes
on a massless spin two closed string excitation,
motivating the natural interpretation of string theory as a
unified formulation of gravity {\em and} gauge theory. The theory
is characterized by a dimensionful scale corresponding to the
string tension, $\alpha^{\prime -1/2}$,
and a dimensionless coupling, $g_s$, the strength of
the cubic string interaction vertex. It is therefore natural
to identify the string tension of fundamental
strings with the Planck mass scale, at which
the gauge and gravitational interactions are of comparable strength.
Note that since closed
string interactions only result in producing closed strings, it is
consistent with perturbative unitarity to have a pure closed string
theory with $g_s$$=$$g_{\rm closed}$.
In the open and closed string theory, consistency requires
in addition the tree level relation between couplings,
$g_{\rm closed}$$=$$g_{\rm open}^2$.

\section{A Weyl-invariant Quantum Mechanics}

\vskip 0.07in
The full beauty of perturbative string theory becomes transparent
upon detailed examination of the loop expansion. The world-sheet representation
of string loop
amplitudes implies the existence of a single graph at each order in
loop perturbation theory. The loop amplitudes in string theory can be
equivalently interpreted as quantum correlators in a two-dimensional
\lq\lq gauge theory", where the gauge symmetry is general coordinate
invariance \cite{polyakov}.
The S-matrix describing the scattering of asymptotic string states is
obtained by invoking two-dimensional conformal invariance to represent
asymptotic on-shell states as operator insertions that are local
in the {\em two-dimensional} sense, i.e., on the worldsheet.
Off-shell string states are
boundaries on the worldsheet--- either macroscopic
closed loops or macroscopic line segments, localized
instead in the embedding {\em spacetime} \cite{cohen}. This correspondence
between worldsheet and spacetime pictures has remarkable consequences.
Consider a gauge invariant path integral expression for the generic
Greens function at arbitrary order in the string loop expansion. Upon
gauge fixing to conformal gauge, the path integral over metrics
is restricted to the fiducial representative from each conformally
inequivalent class of metrics. This is an {\em ordinary} integral over
the finite number of moduli parameters of Riemann surfaces of fixed
topology. Remarkably, in any critical string theory both the measure
of the path integral,
the functional determinants, and vertex operator insertions,
can be unambiguously computed as functions of the moduli, while
preserving the full Diffeomorphism $\times$ Weyl gauge invariance. The
result is an unambiguously normalized and ultraviolet finite expression
for the Greens function with a well-defined, zero string tension, field 
theory limit.

\vskip 0.07in
The resulting expressions for the field theory Greens functions are free
of ultraviolet regulator ambiguity. More importantly, in an infrared
finite string
theory, they are also free of ambiguity in the choice of renormalization
scheme \cite{poltorus,ncom}. Both properties are a consequence
of having maintained manifest two dimensional general coordinate
invariance in computing the full string theory Greens function {\em prior}
to taking the field theory limit defined as follows:
we factorize on massless mode exchange in either open or closed
string sectors, projecting also
onto the massless on- or off-shell modes
in any external states, and integrating out the worldsheet
modulus dependence of the resulting expression.
The result is a coupling in the field theory in
which we can smoothly take the zero string tension limit.
Thus, the expression for any renormalized string theory n-point Greens
function, including in particular the corresponding field theory limit,
is independent of dependence on the string tension, $\alpha^{\prime -1/2}$,
which plays
the role of an ultraviolet cutoff in spacetime.

\vskip 0.07in
String theory can therefore be more simply understood as a
{\em renormalizable scale invariant quantum mechanics} with
a single Wilsonian renormalization in the two-dimensional sense:
the worldsheet cosmological constant renormalizes to zero
in the worldsheet infrared regime, simultaneous with setting the
string tension to zero. Thus, spacetime ultraviolet corresponds to
worldsheet infrared \cite{polbook}.
Corrections to the tree-level relations for the
coupling constants and masses--- measured in units of string tension,
are given at each order in the string loop expansion by an unambiguously
normalized, and finite,
2d general coordinate invariant string amplitude \cite{ncom}.
Factorizing on the lowest-lying modes--- those that are
massless at tree level,
gives the loop corrections to the tree-level values of
the coupling constants and masses in the field
theory. Once again the results are independent of dependence on
the spacetime uv cutoff.
But, more importantly, due to modular invariance of closed string
amplitudes or open-closed world-sheet duality of open and closed
string amplitudes, the worldsheet infrared and
worldsheet ultraviolet
asymptotics of any string amplitude
are closely related \cite{polbook}. A
well-known consequence is the existence of
spacetime UV-IR relations in any perturbative string theory
\cite{polbook,dkps,asymp,bosonic}.

\vskip 0.07in
The renormalizability of perturbative string theory
is obscured from the viewpoint of the spacetime Lagrangian.
Since the renormalized Greens functions for massless fields
are computed in an $\alpha^{\prime}$ expansion using the
nonrenormalizable Wilsonian
effective action with ultraviolet cutoff of order the string scale,
it is simply not possible in this framework to infer
that the cutoff can in fact be removed. The conclusion
that string theory is perturbatively renormalizable relied
crucially on the existence of an all orders in $\alpha^{\prime}$
worldsheet representation of the Greens functions,
recasting the computation as a problem in quantum mechanics
instead of in a field theory.
A remarkable consequence is that, unlike in quantum field theories,
the dimensionless vacuum energy density in string theory---
where we rescale by the spatial volume and inverse temperature
defining, more precisely, the
effective action functional \cite{poltorus,bosonic,fermi},
is {\em finite}: a calculable quantity in any infrared
finite background of the string, as was
first noted in \cite{poltorus}.

\vskip 0.07in
The discovery of supersymmetry in the mass spectrum of
free strings with worldsheet fermionic degrees of freedom
results in supersymmetric string theories, where
nonrenormalization theorems protect the tree-level relations
between masses and couplings. Rapid developments in the eighties
culminated in the discovery of anomaly free superstring theories
with massless fermions in chiral
representations of the gauge group \cite{gsw,polbook}.
Ultraviolet finiteness was firmly established at
the one-loop level for the ten-dimensional SYM-supergravity
obtained in the
low energy field theory limit of vanishing string tension: the
tower of excitations with Planck scale masses
decouples, leaving only massless modes but with the
key non-renormalizable
couplings necessary for establishing
the absence of gauge and gravitational anomalies. The
nonrenormalizable terms are the chief remnant signal in the
ten-dimensional Wilsonian effective Lagrangian of its
string theory origin.

\vskip 0.07in
Discovery of the anomaly-free and UV finite type IB
O(32) theory of open and closed strings, the non-chiral type IIA
and chiral type IIB closed strings, and the heterotic O(32) and
$E_8$$\times$$E_8$ closed strings, brings the number
of distinct perturbatively consistent ten-dimensional supersymmetric
string theories
to five. This puzzling multiplicity of consistent perturbation
theories was finally removed with the insightful union of the
five perturbative string limits by strong-weak coupling and
target space dualities. Finally, with the identification of Dbranes as
nonperturbative gauge-gravity solitons of zero width, charged
also with respect to one or more
antisymmetric p-form potential where $0$ $\le$ $p$
$\le$ $10$,
and playing a crucial role in the conjectured string
dualities, we
have come full circle. We return to discover a new, and deep,
underlying relationship between {\em nonperturbative}
string theory and gauge theory. The focus on reduced matrix
models in \cite{bfkt}, and citations thereof, should be 
understood in light of these previous results.

\section{UV Asymptotics of Open and Closed Strings}

Covariant open and closed string amplitudes are expressed as
reparameterization
invariant sums over {\em open} Riemann surfaces with boundaries.
Surfaces with crosscaps must be included in the Polyakov sum
whenever a Dirichlet boundary is
localized on an {\em orientifold}, an
orientation-reversing hyperplane in the
embedding spacetime \cite{polbook}.
In \cite{asymp} I describe the worldsheet representation of the
generic open and closed string amplitude with Dirichlet boundaries
in the Fenchel-Nielsen parameterization of the moduli space of
Riemann surfaces--- distinguished by both the simplicity of the
measure for moduli, and knowledge of the domain of modular integration
\cite{dhoker}. \cite{asymp} examines the infrared and ultraviolet finiteness
of the generic orientable bosonic open and closed string amplitude
introducing, for the sake of pedagogy, both ultraviolet and infrared regulators
on the worldsheet, preserving both T-duality and open-closed
worldsheet duality. This prescription constitutes an explicit violation of
scale invariance, but it can be verified that the regulators can be
removed at the end of the calculation \cite{dhoker}. \cite{asymp} includes
a self-contained account of a beautiful formalism linking the eigenvalue
problem for the full set of invariant differential operators
on the worldsheet to the geometry of the integration domain of
the moduli of the Riemann surface,
following uniformization to the hyperbolic upper half plane.
The eigenvalue spectrum of the scalar Laplacian has, in turn, a precise
isomorphism to the length spectrum of geodesics,
enabling both deduction of the Fricke-Klein moduli, and an explicit
characterization of the boundary of the modular domain.

\vskip 0.07in
These results enable a rather simple analysis of the spacetime
ultraviolet and infrared asymptotics of generic multiloop amplitudes.
Using Selberg trace techniques \cite{dhoker}, and Huber's exponential bound
on the asymptotic growth in the number of boundary geodesics of fixed
length L, in the limit of large L,
I show conclusively that the spacetime ultraviolet limit is
benign \cite{asymp}, even in open string amplitudes where
the potentially divergent contribution from
the lowest-lying modes of the open string spectrum is included in
the modular integration. Invoking open-closed worldsheet duality,
we can always map this regime into the $l$ $\to$ $0$
regime dominated by the lowest-lying closed string modes.
It can be shown that the potential divergence from
the \lq\lq pinched handle" is given 
by the shortest length geodesic, corresponding
to exchange of a closed string tachyon \cite{dhoker}.
Thus, in the absence of both open and closed string tachyons,
or in an infrared finite string theory, ultraviolet finiteness is
{\em automatic} \cite{polbook}.

\section{UV Asymptotics of Off-shell String Amplitudes}

A manifestly Weyl-invariant Poyakov path integral
representation of off-shell string amplitudes was introduced in
the ingenious paper \cite{cohen}. In \cite{wilsonp}, we formulate
the problem of computing from first principles the pair
correlation function of Wilson loops with fixed spatial
separation in constant
external Maxwell (electric) field. This configuration is a probe
of sub-string-scale short distance physics.
We assume Wilson loops with fixed spatial separation, $r$,
where $r$ is much smaller than the loop length, $l$.
Note that our gauge theory results will
not rely on taking a large N limit,
and are more fruitfully interpreted in the abelian theory
on the worldvolume of each Dbrane, the total number of Dbranes,
N, being fixed by the
overall consistency of the string theory \cite{polbook}.
The Dbrane carries, in addition, a constant external
electric field. In the case of the type I and type II strings,
it may also support one or more constant higher rank antisymmetric
gauge potentials coupling to Dbranes, as in \cite{typeI,flux1}.

\vskip 0.07in
To obtain a gauge theory correlation function leading to
a closed form expression for the short distance sub-string-scale
manifestation of the familiar heavy quark potential, we
extract the worldsheet infrared \lq\lq field theory" asymptotics
of the off-shell closed string propagator between loops, ${\cal C}_{i}$,
${\cal C}_f$, with fixed spatial separation $r$ in spacetime. The loops
represent the proper time evolution of a pair of semiclassical heavy
quarks--- in slow relative motion in a direction, ${\bf{\hat i}}$,
orthogonal to the direction of fixed spatial separation, ${\bf {\hat j}}$.
Thus, the pair of quarks is free to move as a unit in the embedding
spacetime, albeit at small velocity, while maintaining a fixed spatial
separation. In the open and closed bosonic string theory,
the Wilson loops lie in the worldvolume of the spacefilling
D25branes. The small \lq\lq velocity", ${\rm tanh}u^i$, is simply
the strength of the constant electric field, $F^{0i}$, as in
\cite{dkps}.

\vskip 0.07in
Why is this a worldsheet representation of the short distance
Wilson loop correlator, and not that of the short distance
potential between bosonic Dbranes discussed in \cite{dkps}?
The two string computations differ in the specification of the
boundary value problem on the worldsheet.
In \cite{wilsonp}, we sum over all worldsheets of cylindrical
topology connecting two loops with fixed spatial separation,
moving {\em as a unit} within the worldvolume of a
space-filling D25brane. The computation in \cite{dkps}
gives instead the potential between a pair of
D0branes in slow relative
motion, with a slow monotonic increase in spatial separation,
although $r(\tau)$ is assumed to be a sub-string-scale distance.
Wilson loop boundary conditions
in the gauge theory require that we impose Dirichlet boundary
conditions in the string theory, and D25branes are, of course,
hypersurfaces in the embedding spacetime where bosonic open
string endpoints satisfy a Dirichlet boundary condition.
It will become apparent that the
physics is {\em abelian}, independent of the total number of
D25branes which is determined by the overall consistency of
the string theory.

\vskip 0.1in
Consider the classic problem of giving an analytic description of
the short distance sub-string-scale manifestation
of the abelian flux tube linking a pair of semiclassical heavy
quarks. We cannot, of course, extend the effective flux
tube picture of abelian gauge theory into the short distance regime
since the theory is not asymptotically free.
However, in the presence of constant external fields, we can probe
{\em arbitrarily} short
sub-string-scale distances in an open and closed string
theory \cite{dkps,wilsonp,typeI}.
For the bosonic string, where we have no control on strong-weak
coupling duality, we will simply complete the systematics of this worldsheet
calculation ignoring the question of what dynamics stabilizes such a
short abelian flux tube. The crucial new element in the
worldsheet calculation
of the off-shell closed string propagator is
boundary reparameterization invariance.
We complete the work begun in \cite{cohen}, giving a simple
extension of Polyakov's treatment of the manifestly two-dimensional
Diffeomorphism $\times$ Weyl invariant path integral over bulk
worldsheet metrics to the path integral over boundary einbeins,
assuming, for simplicity,
a 1-to-1 mapping of each worldsheet boundary
into a corresponding circular Wilson loop \cite{wilsonp}.

\vskip 0.07in
In the worldsheet infrared limit, and with ${\it l}$ $>>$ $r$,
we obtain an expression for the sub-string-scale manifestation
of the closed time propagator of a flux tube in a nonsupersymmetric
abelian gauge theory.
This is, of course, only a formal solution to the boundary value problem
since we have not addressed the dynamics responsible for the formation
of this classical configuration.
Extracting the effective potential,
$V_{\rm eff} $ $=$ $- \Gamma_0/T V $, where $\Gamma_0$ is
the gauge
theory limit of the off-shell closed string propagator,
we find a universal static potential, $(d-2)/ r$,
where $d$ $=$ $26$ is the dimension of the embedding spacetime.
The velocity dependent
corrections--- present also in the analogous result for
the type IIB superstring, can be computed in a systematic double expansion
in small velocity and short distance. The result probes distance
scales down to $r_{\rm min}^2$$\sim$$ 2\pi\alpha^{\prime} {\rm tanh}^{-1} v$,
assuming nonrelativistic velocity $v$.

\vskip 0.07in
Ref.\ \cite{typeI} gives the supersymmetrization of this result,
summing over surfaces of cylindrical topology between closely separated
circular Wilson loops lying in the
worldvolume of a spacefilling Dbrane in the type IIB superstring.
The phases in the path integral representation of
the off-shell type IIB closed string propagator between Wilson loops
are, a priori, unknown, and correspond to
the weighted sum over worldsheet spin structures.
We will require the absence of both the leading contribution, which is 
a tachyonic mode, and the next-to-leading term, which gives the static
potential. This ensures that upon application of open-closed
worldsheet duality--- which maps the short distance gauge potential
into a long distance supergravity potential \cite{dkps,polbook},
we preserve the full N=2 spacetime supersymmetry
for vanishing constant external field.
Turning on a constant electric field induces an
external field-dependent short distance potential, also breaking half of
the spacetime supersymmetries. For Wilson loops, ${\cal C}_i$, ${\cal C}_f$,
wrapped
about some spacelike compact direction the potential
takes the form: $V(r) $$=$$ 2^4 \pi^{7/2} \alpha^{\prime 4} \Gamma(9/2) \left
( {{({\rm tanh}^{-1} v)^4}\over{r^9}} \right )$.

\vskip 0.07in
A puzzle addressed in \cite{flux1} is to explain what accounts for the
stability of the short abelian flux tube. Consider a configuration of
closely separated parallel IB soliton strings: D1branes, with worldvolume
quantized constant $C_0$ and $B_{\mu\nu}$ potential, wrapped about the compact
$X^9$ coordinate and with fixed spatial separation in the $X^8$ direction.
They move as a unit within the worldvolume of a space-filling D9brane under
the action of an external electric field. In nine dimensions, the type IB
$O(32)$ string is the strong coupling dual of the heterotic $SO(32)$ string,
with identical non-abelian gauge group. Thus, this is nothing but the strong
coupling limit of a pair of closely separated wrapped heterotic soliton strings,
in a nonperturbative background characterized by both Yang-Mills gauge fields
and discrete moduli. A nice check is to consider the inverse of the
orientation reversal projection, arriving at a configuration of localized
parallel wrapped soliton strings, in the absence of an external electric field,
in the 9d IIB massive supergravity \cite{hull}. The Dstrings couple to
the Ramond-Ramond $*F_{10}$ scalar background field strength. A T$^9$-duality
gives a pair of D0branes in the worldvolume of the D8brane stack with
half-integer quantized $B_{NS-NS}$ field strength. The radius $R_{9B}$
corresponds to the mass parameter of the massive IIA supergravity
\cite{hull}, which is quantized in integer units of the inverse IIB radius.

\vskip 0.07in
The short tube of abelian flux runs between D0brane sources with fixed spatial
separation. It is amusing that this configuration fits both the geometry,
and quantized constant background fields, of a well-known massive string
soliton of the nine-dimensional IIA supergravity \cite{hull}. In a constant
external electric field, $F^{09}$,
a pair of D8branes, carrying a D0brane and its image brane,
will be separated by a sub-string-scale
distance in a spatial direction orthogonal to the direction of
nearest separation, which is held fixed, and which lies
within the worldvolume of the D8brane.
The difference between computing the potential between a pair of
D0brane sources moving as a unit within the worldvolume of the
embedding D8brane, and that between a pair of D0branes in slow relative
motion with respect to each other is simple: the $r^{-7}$ falloff
computed in \cite{dkps} is replaced by the more rapid $r^{-9}$ falloff,
characterizing the {\em mass} of the bound state of D0branes
moving as a unit under the influence of the external electric field.
Note that the Wilson loops, ${\cal C}_i$,
${\cal C}_f$, are a fixed distance $r$ apart, but there is no constraint on
their {\em position}. Thus, we have a zero mode corresponding
to the center of mass motion of the D0brane pair, parallel to their
direction of nearest separation.

\vskip 0.07in
What if we take the strong coupling limit of the type I$^{\prime}$
theory assuming the nonabelian gauge group $E_8$$\times$$E_8$? The result
can be interpreted as a membrane of finite sub-string-scale width in M
theory stretched between 10d walls at the endpoints of the finite interval
$X^{10}$, the strong coupling limit of the heterotic $E_8$$\times$$E_8$
string.

\section{String Theory in Noncommutative Spacetime}

The manifestly Weyl invariant path integral computation
of the closed string tachyon scattering amplitude at one-loop 
given in \cite{poltorus}, is extended to 
an analogous
result for the scattering of open string tachyons on the
boundaries of both the planar, and nonplanar, cylinder amplitude in
bosonic string theory, and in the presence of a
constant antisymmetric tensor two-form potential, in \cite{ncom}. The
worldvolume of the Dbrane coupled to a background two-form
potential is a noncommutative spacetime \cite{ncom}.
The string amplitudes are nevertheless found to be both
finite and unambiguously normalized, including in
the limit of zero string tension. We infer, by similar arguments as
given for vanishing external field,
the exact Wilsonian renormalizability of open and closed string
theory on a noncommutative space. The ultraviolet cutoff can be taken to
infinity. The ordinary integral over the cylinder modulus can,
in fact, be carried out in closed form upon taking the
worldsheet infrared limit of the expression for the planar amplitude.
We demonstrate that the momentum dependent phase factor in the
one-loop noncommutative field theory amplitude can indeed be interpreted
as a wave-function renormalization, and we obtain its explicit
form in terms of the star product. Coupling constant renormalization is exactly
reminiscent of that in commutative spacetime, with identical short
distance singularity but for a finite renormalization of the open
string mass scale: $(2\pi {\alpha}^{\prime})^{-(p+1)/2} {\rm det} ({\bf 1}+
{\bf B})$, where for simplicity, we assume a Dpbrane with $p$ odd, and
$\le$ $25$. The higher effective string tension implies that
open string modes probe shorter distances than the closed string scale.

\vskip 0.07in
The opposite regime of zero momentum transfer is dominated by the lightest
closed string modes, or the worldsheet ultraviolet asymptotics.
We will find that the zero momentum transfer limit of the nonplanar
amplitude is ordinary gravity with closed string masses scaling
in units of the bare string tension, $(2\pi \alpha')^{-1/2}$.
Thus, the infrared regime is completely benign and the puzzling
inconsistencies of the infrared regime present in noncommutative field
theory are circumvented.

\section{Type I Duals of the CHL Models}

In the appendix of \cite{flux1}, I gave an
analysis of enhanced gauge symmetry and nonperturbative type I$^{\prime}$
states using an extended IIB-IB-I$^{\prime}$ chain in conjunction with S- and
T- dualities. The argument is as follows. The states in the 
spinor representation of $O(16)$ necessary 
to obtain the 9D IB dual of the $E_8$$\times$$E_8$ enhanced
symmetry point in the moduli space of the 9D heterotic string requires
{\em nonperturbative} IB states.
Consider the nonperturbative states associated with
a pair of D1branes wrapped in the $X^9$ direction and lying in the
worldvolume of D9branes carrying $O(16)$$\times$$O(16)$ Chan-Paton
factors. A $T_9$ duality maps this into a type I$^{\prime}$ background, with a
D0brane lying in the worldvolume of the stack of 8 D8branes on either of two
orientifold planes. The spinor states are associated with configurations of
eight D1-D9 strings, or their $T_9$-dual
D0-D8 strings, at each of two orientifold planes.
The counting of extra massless modes is given by the $(SU(2))^8$ decomposition
of the vector and spinor representations of $SO(16)$. The D1-D9 strings live in
doublets of the eight $SU(2)$'s, and the projection to the spinor and conjugate
spinor is isomorphic to the heterotic weight lattice of $E_8$. It will be of
great interest to give a more detailed description of such nonperturbative
excitations of the D1-D9 IB configuration.

\vskip 0.07in
At a special radius of $X^9$ there is an enhancement of the gauge symmetry to an
$SU(2)$. As noted in \cite{pw}, the IB string coupling diverges
as the heterotic string approaches its self-dual radius. 
Nevertheless, the multiplicity
of additional massless IB states can be inferred by mapping IB to IIB by
an inverse orientifold transformation, and then using S-duality to map the
wrapped Dstrings to wrapped F-strings: at its self-dual 
radius, the F-string winding
states give the well-known enhancement of $U(1)$ to a full $SU(2)$.
This argument relies on special properties of theories with 16 supercharges:
S-duality
is well-established, and orientation reversal is a freely acting $Z_2$ symmetry.
With these two extensions to the moduli space of IB backgrounds with 16
supercharges, we can identify IB strong coupling duals for all of the heterotic
CHL orbifolds \cite{chl}. This is remarkable and striking confirmation of the
validity of type I-heterotic duality \cite{pw}.

\section{The Free Closed String Gas}

In \cite{us} I gave a world-sheet representation of the pair correlation
function of closely separated time-like Wilson loops, an order parameter for
a phase transition to the long string phase in the type I open and closed
string gas described more completely in \cite{fermi}. In formulating the 
correct set-up for this calculation I
became aware of several puzzles and loopholes in the standard treatment of the 
simpler case of closed string thermodynamics. Let us begin
with a gas of free closed bosonic strings, computing the effective
action functional at one-loop at finite temperature $\beta$$=$$1/T$. In the
sequence of papers \cite{bosonic,fermi,us}, we establish several points of
interest that correct misconceptions in the standard treatment. Our starting 
point for
the free energy of the closed string gas is the modular invariant expression
obtained directly from the effective action functional: 
$F$ $=$ $-W/\beta$,
where $W$ is the sum over connected vacuum string graphs at finite temperature.
$W$ is an intensive thermodynamic variable. The world-sheet representation 
of $W$ is the Polyakov path integral without vertex operator insertions.

\vskip 0.07in
An important point of interest is the absence of a Hagedorn phase transition in the
free energy of a closed string gas in the absence of tachyonic modes. This 
cannot be illustrated in the pedagogical, but also unphysical, case of the 
bosonic string gas since the string spectrum contains a zero temperature 
tachyon. In the pedagogically similar, but physically meaningful, heterotic 
string gas, it is easy to 
establish that the free energy is a finite and normalizable function at all
temperatures starting from zero \cite{fermi}. 
The reason is modular invariance. In particular, notice that quite apart
from the occurence of tachyonic winding modes at high temperature the fermionic 
closed string gas also has potential tachyonic momentum modes at low temperature.
Since there is no tachyonic mode in a supersymmetric fermionic string gas, whether
type IIA, IIB, or heterotic, at zero temperature, it would be unphysical to have
a tachyon at infinitesimal temperature. In \cite{fermi}, we show that in the 
absence of Ramond-Ramond backgrounds there are no tachyon-free solutions for a 
modular invariant one-loop vacuum amplitude at finite temperature in the type 
II case, which also recover 
the correct zero temperature limit. On the other hand, in the heterotic 
string gas, in 
the presence of a temperature-dependent Wilson line background we are able to
successfully suppress the appearance of tachyons at all temperatures starting 
from zero. The heterotic string gas has no exponential divergences in the free
energy and no Hagedorn phase transition. Instead, it displays a self-dual phase
transition at $T_C$$=$$1/\pi \alpha^{\prime 1/2}$ belonging to the universality class of
the Kosterlitz-Thouless phase transition. The transition is in every respect
similar to that exhibited by the free closed bosonic string gas, except that
the unphysical fixed point entropy that appears for the bosonic string gas 
\cite{bosonic} is absent in the heterotic case. The free energy, and all of its 
partial derivatives with respect to temperature, are continuous through the 
transition. The type II string gases exhibit a similar phase transition: type IIA
is mapped to type IIB. The IIA winding modes are interchanged with IIB momentum
modes, and vice versa \cite{fermi}.

\vskip 0.07in
In summary, for the infrared-finite heterotic string gas with monotonically
increasing internal energy and positive specific heat at criticality, we obtain a
thermodynamically stable ensemble at all temperatures starting from zero. The 
nonabelian
gauge group is $SO(16)$$\times$$SO(16)$. We should note that the analogous 
Lorentzian time nonsupersymmetric heterotic string ground state was 
discovered in \cite{agm}.
Recall that the finite temperature string gas is
defined in the presence 
of a temperature-dependent self-dual background, ${\bf A}_0 (\beta)$$=$$1/\beta$. 
In terms 
of the low energy gauge theory limit, this
corresponds to a modification of the usual axial gauge quantization. Here,
$A_0$ has been set to a temperature dependent constant rather than zero.

\vskip 0.07in
We emphasize that gauge fields appear to be essential in order that a given 
supersymmetric ground state of string theory have a sensible finite temperature
description. We have shown that the type IIA and IIB ten-dimensional string gas 
contains tachyonic modes at any infinitesimal temperature different from zero. 
A plausible resolution is discussed in \cite{fermi}. Of necessity, it requires
a nontrivial Ramond-Ramond sector and nonperturbative gauge fields.

\section{The Transition to the Long String Phase}

The open and closed string theory can be mapped into its thermal duality
transform, the type ${\tilde{I}}$ string containing thermal Dpbranes: extended
objects with a $p$-dimensional spatial worldvolume \cite{us}. The low energy 
field theory
on the worldvolume of a thermal Dpbrane is a $p$-dimensional finite temperature
Higgs-gauge-gravity theory. As shown in \cite{fermi}, the ten-dimensional 
type IB theory also has a stable
finite temperature ground state characterized by the nonabelian gauge group
$SO(16)$$\times$$SO(16)$, and a tachyon-free spectrum at all temperatures
different from zero. The amplitude is tadpole-free, and the one-loop free 
energy is found to vanish identically \cite{fermi}. We comment that this 
result is consistent with the weak-strong coupling, heterotic-type I, duality
conjectures.
This is not too surprising since, at least for infinitesimal
temperatures different from zero, the strong-weak duality
relations would be expected to hold with small $\beta$-dependent modifications.
It is not implausible that the heterotic ground state at finite temperature
with finite oneloop vacuum cosmological constant and, consequently, a 
dilaton tadpole, {\em flows} to strong coupling. This strongly coupled 
heterotic ground state has a weakly-coupled type I description as given 
in \cite{fermi}. While satisfactory from a physical standpoint, these inferences
must remain conjectural in the absence of a nonperturbative formalism. 

\vskip 0.07in
It is natural to look for an analog of the deconfinement phase transition 
of gauge theories in our string theory calculations. Accordingly, we compute 
the pair correlator of parallel and closely separated timelike Wilson loops 
lying in a thermal Dpbrane. Isolating the low temperature behavior of the 
gas of short open strings, we find a $T^{10}$ growth characteristic of the 
ten-dimensional finite temperature gauge theory characterizing this limit. 
At a temperature
at, or below, $T_C$ we observe a transition to a long string phase 
characterized by a pair potential approaching a constant. 
On the other hand, below $T_C$,
we can expand in a Taylor expansion to obtain a $1/r^3$ correction to the 
leading $1/r$ dependence. 
Precisely at the transition, 
we observe 
an inverse power law fall-off with a coefficent independent of the 
dimensionality of the Dpbrane in question \cite{us,fermi}. 
In the presence of an external electric field the transition temperature is
modified with the simple replacement, $T_C$$\to$$uT_C$, where 
$u$$=$${\rm tanh}^{-1} {\cal F}_{09}$ \cite{fermi}, where ${\cal F}$ is
the electric field strength \cite{us,fermi}.  

\section{Conclusions}

\vskip 0.1in
I have emphasized at the outset the importance of refining our understanding
of
the details of the worldsheet prescription for string scattering amplitudes
because of the insight it gives into the fully nonperturbative string/M theory.
In particular, the world-sheet representation of off-shell string amplitudes
is largely unexplored territory, with the potential of suggesting many new
worldsheet calculations with field theory limit results that are
of significance in both gauge and gravitational physics. Extension 
of the worldsheet representation to off-shell amplitudes with 
finite-sized boundary segments is of interest. The inclusion
of non-trivial wavefunctions on the boundary representing physical
semiclassical D0branes 
could be of far-reaching importance, raising the possibility of a 
first-principles computation of both the phase and normalization of 
the \lq\lq meson" and \lq\lq hadron" D0brane bound state wavefunctions.
Extended exploration of both strong-weak coupling and target spacetime dualities
in the context of off-shell string amplitudes, along the lines of the finite-width
bound states described in section IV, remains a goal for future work.
In particular, it could facilitate understanding the precise role
of the NS spacetime solitons in nonperturbative string/M theory.
Finally, the subject of string thermodynamics has far to go. We have
exhaustively studied the free string limit. If feasible, an extension 
of the worldsheet representation that describes non-equilibrium string 
thermodynamics would be of great interest. A fully nonperturbative
formalism that goes beyond the limitations of the worldsheet remains 
a goal for future work.

\vskip 0.2in
\noindent{\bf ACKNOWLEDGMENTS}:
I thank the symposium organizers, especially Tada-san, for the
invitation to present this work and the participants for their interest.
I would like to acknowledge the hospitality of Prof.\ H.\ Kawai, the Physics
Department of Kyoto University, and the
Yukawa Institute for Theoretical Physics, during completion of this work.
I am especially grateful to Kawai-san for many stimulating discussions.


\begin{thebibliography}{99}
\bibitem{bfkt}T. Banks, W. Fischler, S. H. Shenker and L. Susskind,
Phys. Rev. D55 (1997) 5112. N. Ishibashi, 
H. Kawai, Y. Kitazawa, and A. Tsuchiya,
Nucl. Phys. {\bf B510} (1998) 158.
S.\ Chaudhuri, hep-th/0202138.
\bibitem{polyakov} A. M. Polyakov,
Phys.\ Lett.\ {\bf B103} 207 (1980).
\bibitem{poltorus}J. Polchinski,
Comm. Math. Phys. {\bf 104} (1986) 37.
\bibitem{gsw}M. Green, J. Schwarz, and E. Witten, {\it String Theory},
Volumes I \& II (Cambridge) 1987.
\bibitem{polbook} J. Polchinski, {\it String Theory},
Volumes I \& II (Cambridge) 1998. See, Ch.\ 9.5, 13.5, 8.2, and 9.8.
\bibitem{pw}J. Polchinski and E. Witten, Nucl.\ Phys.\
{\bf B460} (1996) 525.
\bibitem{dhoker}
A. Selberg, J. Indian Math. Soc. Vol. 20, 47 (1956).
H. P. McKean, Comm. Pure \& Applied Math., Vol. XXV, 225 (1972).
S. Wolpert, Ann. Math. {\bf 169} (1969) 323;
Comm. Math. Phys.  112 (1987) 283.
E. D'Hoker and D.-H. Phong, Nucl. Phys. {\bf B269} (1986) 205.
\bibitem{asymp}
L. Bers, Bull. London Math. Soc. 4 (1972) 257.
L. Keen, Annals Math. Studies, Vol. 66 (1971) 205.
H. Huber, Math.\ Annals 138 (1959) 1.
S. Chaudhuri, JHEP 003 (1999) 008.
\bibitem{ncom}S. Chaudhuri and E. Novak, 
JHEP 008 (2000) 027.
\bibitem{cohen} A. Cohen, G. Moore, P. Nelson, and J. Polchinski,
Nucl. Phys. {\bf B267} 143 (1986).
\bibitem{wilsonp}S. Chaudhuri, Y. Chen, and E. Novak,
Phys. Rev. {\bf D62} (2000) 026004.
\bibitem{typeI} S. Chaudhuri and E. Novak, Phys.\ Rev.\ {\bf D62}
(2000) 046002.
\bibitem{dkps}M. Douglas, D. Kabat, P. Pouliot, and S. Shenker,
Nucl.\ Phys.\ {\bf B485} (1997) 85.
C. P. Bachas, Phys.\ Lett.\ {\bf B374} (1996) 37.
\bibitem{hull}E. Bergshoeff, M. de Roo, M. Green, G. Papadopoulos, P. Townsend,
Nucl.\ Phys.\ {\bf B470} 113. C. Hull, JHEP 9811 (1998) 027.
B. Janssen, P. Meessen, and T. Ortin,
Phys.\ Lett.\ {\bf B453} (1999) 229. M. Massar and J. Troost,
Phys.\ Lett.\ {\bf B458} (1999) 283.
\bibitem{flux1}S.\ Chaudhuri, Nucl.\ Phys.\ {\bf B591} (2000)
243.
\bibitem{chl} S. Chaudhuri, G. Hockney, and J. Lykken,
Phys. Rev.\ Lett.\ {\bf 75} (1995) 2264. S. Chaudhuri and J. Polchinski,
Phys.\ Rev.\ {\bf D52} (1995) 7168.
A. Mikhailov, Nucl.\ Phys.\ {\bf B534} (1998) 612.
\bibitem{hagedorn} R. Hagedorn, Nuovo Cim.\ Suppl.\ 3 (1965) 147.
G. Hardy and S. Ramanujan, Proc.\ Lond.\ Math.\ Soc.\
{\bf 17} 75 (1918). K. Huang and S. Weinberg,
Phys.\ Rev.\ Lett.\ {\bf 25} (1970) 895.
P. Salomonson and B. Skagerstam,
Nucl.\ Phys.\ {\bf B268} (1986) 349.
J. Atick and E. Witten,
Nucl.\ Phys.\ {\bf B310} (1988) 291.
\bibitem{mclain}B. Maclain and B. Roth, Comm.\ Math.\ Phys.\
{\bf 111} (1987) 1184. K. O'Brien and Chung-I Tan,
Phys.\ Rev.\ {\bf D36} (1987) 1184.
\bibitem{ckt}S.\ Chaudhuri, H.\ Kawai, and S.-H.\ H.\ Tye,
Phys.\ Rev.\ {\bf D36} 1148 (1987).
H. Kawai, D. Lewellen, and S.-H.H. Tye,
Nucl.\ Phys.\ {\bf B288} 1 (1987).
\bibitem{gin}P. Ginsparg,
Nucl.\ Phys.\ {\bf B295} (1988) 153.
\bibitem{kosterlitz}J.\ M.\ Kosterlitz,
J.\ Phys.\ {\bf C7} (1974) 1046.
\bibitem{agm}N. Seiberg and E. Witten, Nucl.\ Phys.\ {\bf B276} (1986) 272.
L. Alvarez-Gaume, P. Ginsparg, G. Moore, and C. Vafa,
Phys.\ Lett.\ {\bf B171} (1986) 155. L. Dixon
and J. Harvey, Nucl.\ Phys.\ {\bf B274} 93 (1986). H. Kawai,
D. Lewellen, and S.-H. H. Tye, Phys.\ Rev.\ {\bf D34} (1986) 3794.
\bibitem{us} S. Chaudhuri,
Phys.\ Rev.\ Lett.\ {\bf 86}, 1943 (2001),
hep-th/0008131. 
\bibitem{bosonic}S.\ Chaudhuri, Phys.\ Rev.\ {\bf D65} (2002) 066008, 
hep-th/0105110.
\bibitem{fermi} S. Chaudhuri, hep-th/0208112. See, also, the related 
discussion in hep-th/0203058.
\end{thebibliography}
\end{document}